\documentclass[twocolumn,showpacs,preprintnumbers,amsmath,amssymb]{revtex4}

\input {epsf.sty}
\usepackage{graphicx}% Include figure file
\usepackage{epsfig}
\newcommand{\be}[1]{ \begin{eqnarray} \mbox{$\label{#1}$} }

\newcommand{\ee}{\end{eqnarray}}

\newcounter{mycount}

\newcommand\ie {{\it i.e. }}
\newcommand\eg {{\it e.g. }}

\begin{document}

\title{Condensing non-Abelian quasiparticles }
\author{M. Hermanns} 
\affiliation {Department of Physics, Stockholm University
AlbaNova University Center,
SE - 106 91 Stockholm, Sweden}

\date{\today}

\begin{abstract} 
A most interesting feature of certain fractional quantum Hall states is that their quasiparticles 
obey non-Abelian  fractional statistics.
 So far, candidate non-Abelian wave functions have been constructed from conformal blocks in cleverly chosen conformal field theories. In this work we present a hierarchy scheme by which we can construct daughter states by condensing non-Abelian quasiparticles (as opposed to quasiholes) in a parent state, and show that the daughters have a  non-Abelian statistics that differ from the parent. 
In particular, we discuss the daughter of the bosonic, spin-polarized Moore-Read state at $\nu=4/3$ as an explicit example.

\end{abstract}

\maketitle

\newcommand{\etab}{\bar\eta}
\newcommand{\zbar}{\bar z}
\newcommand{\tvphi}{\tilde\varphi}
\newcommand{\rmd}{\mathrm d}
\newcommand{\pplus}{\mathcal{P}_+}
\newcommand{\pmin}{\mathcal{P}_-}
\newcommand{\pone}{\mathcal{P}_1}
\newcommand{\ptwo}{\mathcal{P}_2}

In recent years the quantum Hall (QH) effect has regained much attention because of the exciting possibility of having non-Abelian (NA) excitations, which can in principle be used to construct topologically protected qubits, see \eg Ref. \cite{qbits} for a review. 

The QH states in the lowest Landau level (LLL), which are all believed to be  Abelian, are very well understood in terms of two, partly equivalent schemes. 
The Haldane-Halperin(HH) hierarchy\cite{halphi,haldhi} 
constructs QH states at level $n+1$ as coherent superpositions of quasihole or quasiparticle excitations in a level $n$  parent state. Starting with the Laughlin wave functions at filling fractions $\nu=1/m$, $m$ odd, one can build a hierarchy of daughter states, including all observed QH states in the LLL. The resulting wave functions, however, involve multiple integrals over all quasiparticle positions, which cannot, in general, be evaluated analytically.  Another approach is the composite fermion picture of Jain\cite{jain}, which postulates the existence of composite fermions - flux quanta bound to an electron - that fill effective Landau levels in a reduced magnetic field. The wave function is then obtained by projecting this state to the LLL. In contrast to the hierarchy scheme it yields explicit wave functions, but it does not explain all the QH fractions observed in the LLL. These two approaches are not exclusive - in fact, it has been shown that some Jain states can explicitly be written in a hierarchical form\cite{qeop}. 

In the second Landau level there is no general approach that can explain all the observed fractions. Instead one tries to identify them one by one with candidate states constructed using  conformal field theories (CFTs). 
The connection between QH states and CFTs dates back to the observation that the Laughlin wave functions can be written as  CFT correlators of free, massless bosons\cite{fubini}, and to the discovery of the relation between effective Chern-Simons theories and the conformal blocks of certain CFTs\cite{witten}.
Moore and Read proposed that the QH - CFT connection is very general and used the CFT of the Ising model to construct the  now famous Moore-Read (MR) wave function which is a candidate for the QH state observed  at filling fraction $\nu=5/2$\cite{mr}. Most interestingly, they also showed that the quasiholes in this state obey NA statistics and $su(2)_2$ (or Ising) type fusion rules.  
Later, other CFTs describing parafermions were used to construct the Read-Rezayi (RR) state\cite{RR}, which provide a candidate for the observed state at $12/5$.

In spite of this success, there is no general scheme similar to the successful hierarchy construction of the Abelian states in the LLL. At first, such an approach looks prohibitively difficult, since it would require the explicit construction of many-particle wave  functions for NA particles. In this paper we show how to overcome this difficulty by applying the methods developed in Ref. \onlinecite{qeop} to construct states of many NA quasiparticles and explicitly carrying out the integrations over their positions to obtain a closed expression for the condensate daughter states.  
Using the MR state as an example, we construct explicit trial wave functions for the daughter states and show that they have a NA structure that differs from that of the parent state. 
In the simplest case, the bosonic daughter state at $\nu=4/3$ has quasiholes obeying $su(3)_2$  fusion rules, while the parent bosonic MR state at $\nu = 1$ has $su(2)_2$ type quasiparticles. In this particular case, we are also able to fully identify the relevant CFT and explicitly express our condensate wave function as sums of conformal blocks. For more general NA condensate states, this is not possible, but using techniques developed in Ref. \onlinecite{tt_fusion} we can nevertheless determine the NA fusion rules. 
Although we use the MR state as an example, our method generalizes naturally to other NA states, such as those in the RR series. 
Note that a hierarchy scheme based on condensing {\em Abelian} quasiholes and quasiparticles was proposed earlier by Bonderson and Slingerland\cite{BS}, yielding hierarchical wave functions with the same NA structure as the parent state.

Originally, the MR state, and its quasihole excitations, were written as conformal blocks of correlators involving operators of the Ising CFT. 
In this formulation, the NA statistics is manifest in the monodromies of the conformal blocks representing the  quasihole wave function\cite{naywil}.
This is based on the assumption that there are no extra Berry phases related to quasiparticle braiding, which is supported by analytical arguments \cite{read09} and numerical calculations\cite{NAnum}. It is, however, possible to shift the NA statistics between the monodromies and the Berry phases, and in the following it will be 
important  to use a description where the statistics resides completely in the latter and the monodromies are trivial\cite{qeop}. 
Therefore, we use an alternative description of the MR state as an (anti)symmetrized two-layer state\cite{cappelli}. 
The particle operators-- describing bosons (electrons) for $q$ odd (even)-- are given by $V(z)=e^{i\sqrt{q}\varphi(z)}\cos(\phi(z))$,  where $\varphi$ and $\phi$ are independent, free, chiral bosonic fields with the Greens function $\langle \varphi(z)\varphi(w)\rangle=-\ln (z-w)$.
Here, $\phi$ refers to the layer, while $\varphi$ is related to the electric charge. The ground state wave function is obtained by computing the CFT correlator of $N$ such operators,  $\Psi_{gs}=\langle \prod_{j=1}^{N} V(z_j) \rangle = \mbox{Pf}(\frac{1}{z_i-z_j}) \prod_{i<j}(z_i-z_j)^q $.
The NA quasihole is represented by two operators $H_\pm=e^{i/2\sqrt 2 \varphi(\eta)\pm i/2 \phi(\eta)}$, each corresponding to a hole in one of the layers, and the topological degeneracy of the multi-quasihole state is encoded in the distinct orderings of $H_+$ and $H_-$. By using an identity derived in Ref. \cite{naywil}, one can verify that for $2n$ quasiholes at fixed positions there are indeed $2^{n-1}$ degenerate states as is expected from the CFT analysis.

For the quasiparticle excitations we use the quasilocal operator $\mathcal{P}(\eta)$, introduced in Ref. \cite{qeop}. 
In this context, it is sufficient to know that there are two such operators, $\mathcal{P}_\pm$, corresponding to the two quasiholes.  
Successive insertion of pairs $\pplus \pmin$ into the correlator of the ground state $\Psi_{gs}$ yields the multi-quasiparticle states, for example:
\begin{multline}\label{multiqp}
\Psi (\{z_i\} ; \{\eta_i\})=\langle \prod_{j=1}^{M}\pplus(\eta_j) \prod_{j=M+1}^{2M}\pmin(\eta_{j})\prod_{j=1}^N V(z_j)\rangle\\
= \sum_{\sigma\in S_{N}} (-1)^\sigma e^{-\frac{q_h}{4\ell^2} \sum_{j=1}^{2M}(|\eta_j|^2-2\etab_j z_{\sigma_j})} 
\\ \times 
\langle \prod_{j=1}^{M}V'_+(z_{\sigma_j})\prod_{j=M+1}^{2M}V'_-(z_{\sigma_j})\prod_{j=2M+1}^{N}V(z_{\sigma_j}) \rangle ,
\end{multline}
where $\sigma_j=\sigma(j)$, and $(-1)^\sigma=-1$ for $q$ even and an odd permutation $\sigma$, and $+1$ otherwise.  
We also introduced new particle operators $V'_\pm=\partial_z e^{i\frac{2q-1}{\sqrt{4q}}\varphi(z)\pm \frac{i}{2} \phi(z)+i\sqrt{\frac{2q-1}{4q}}\chi_1(z)\pm \frac{i}{2} \chi_2(z)}$, with $\chi_1$ and $\chi_2$ being independent bosonic fields with the same normalization as $\varphi$ and $\phi$. It is straightforward to compute the correlator in \eqref{multiqp}, to get a product of Jastrow factors acted upon by holomorphic derivatives.

Other candidate wave functions are obtained by considering different orderings of  $\pplus$ and $\pmin$, which is equivalent to permuting the $\eta_j$'s in the exponential in \eqref{multiqp}. The different orderings yield distinct, but not necessarily linearly independent wave functions. Similar arguments as for the quasiholes show that the multiplicity of the $2n$ quasiparticle state is again $2^{n-1}$. 
As already mentioned,  the manifest monodromies in expression \eqref{multiqp} for braiding quasiparticles is zero, and the statistical phase is completely in the Berry phase.

Now we show how to construct hierarchical states from these NA quasiparticles. The construction parallels that of writing the $\nu=2/5$ Jain state explicitly as a quasiparticle condensate in the $\nu=1/3$ Laughlin state\cite{qeop}. According to the original proposal, a state at hierarchy level $n+1$ can be written as \cite{halphi},
%a
%
\be{hiwf}\label{HHhier}
\Psi_{n+1} =   \int  d^2\eta_1 \dots   \int  d^2\eta_{2M}    \, \Phi^\star(\{\eta_i\} ) \Psi_n (\{z_i\} ; \{\eta_i\}) \, ,
\ee
where $\Phi^\star $ is a suitably chosen  "pseudo wave function" for the strongly correlated state formed by the quasiparticles when condensing, and $\Psi_n$ is a multi-quasiparticle wave function of the type \eqref{multiqp}. Even though for fixed $M$ there are in general many linearly independent wave functions, we need only consider one of them, as all orderings yield the same condensate state. This is due to the integration over the $\eta$'s, which amounts to averaging over the positions, and thus also averages  over all possible orderings of $\pplus$ and $\pmin$. Note that for the integrals in \eqref{HHhier} to be well defined, it is crucial to use a $\Psi_n (\{z_i\} ; \{\eta_i\})$ with trivial monodromies in the quasiparticle positions. 

The appropriate pseudo wave functions, $\Phi^\star$ are straightforward generalizations of the ones used in the  Abelian hierarchy.  There are two physical constraints that must be fulfilled. First $\Phi^\star$ must reflect the manifest monodromies of the quasiparticles, \ie it must be totally antisymmetric in $I_+$ and $I_-$ respectively, where $I_\pm$ is the set of $\mathcal{P}_\pm$ coordinates. In addition, homogeneity determines the total degree of the polynomials in  the $\eta$'s (up to terms of order $\mathcal{O}(M)$).  The most natural choice for $\Phi^\star$ is the one appropriate to describe two types of charge $-q_h$ fermions in the  LLL, $\Phi^\star(\{\eta_i \} )=\prod_{{i<j\in I_+}}(\eta_i-\eta_j)^{2p-1}\prod_{{a<b\in I_-}}(\eta_a-\eta_b)^{2p-1} e^{-\frac{q_h}{4\ell^2}\sum_{j=1}^{2M}|\eta_j|^2}$, where $p\geq 1$ is related to the density of the quasiparticle droplet. 
Note that this ansatz, albeit being the simplest and most natural, is not  unique. For instance, one can multiply $\Phi^\star$ by a symmetric polynomial of order $\leq M$ without violating the requirements of antisymmetry and homogeneity. 

In contrast to the corresponding expressions for quasihole condensates, the integrals in  \eqref{HHhier} can be evaluated analytically, as the exponential factors combine to a LLL delta-function \cite{qeop}, giving
\begin{align}\label{NAcond}
 \Psi_\nu&=\sum_{\sigma\in S_{N}} (-1)^\sigma \prod_{{i<j}\atop{i,j=1}}^{M}(z_{\sigma_i}-z_{\sigma_j})^{2p-1}\prod_{{i<j}\atop {i,j=M+1}}^{2M}(z_{\sigma_i}-z_{\sigma_j})^{2p-1}
 \nonumber\\ &\times 
\langle \prod_{j=1}^{M}V'_+(z_{\sigma_j})\prod_{j=M+1}^{2M}V'_-(z_{\sigma_j})\prod_{j=2M+1}^{N}V(z_{\sigma_j}) \rangle,
\end{align}
in the same notation as above. 
The state \eqref{NAcond} has  filling fraction $\nu=4p/(4pq-1)$, and the number of operators is fixed to  $M=N/4p$ by requiring homogeneity.  These states have fundamental quasiholes with electric charge $e/(4pq-1)$ and $\mathbb{Z}_{4p-1}$ parafermion type fusion rules in the sense that the Bratteli diagram is identical to that of the $\mathbb{Z}_{4p-1}$ parafermion CFT. 
Before proceeding to give the arguments for these assertions, some general comments are necessary.  

We mentioned already that the bosonic MR state  can be regarded as a symmetrized two-layer state, with filling fraction $\nu=1/2$ in each of the completely independent layers. The same interpretation is possible for the bosonic ($q=1$) states \eqref{NAcond}, but with a bosonic Jain state instead of a bosonic Laughlin state in each of the layers. 
With this interpretation at hand, the states \eqref{NAcond}  appear as a rather natural and simple generalization of the MR state. However, we want to strongly emphasize  that \eqref{NAcond} is obtained by a well-defined procedure with a simple physical interpretation, namely the condensation of the fundamental NA quasiparticles. As already mentioned, this  is a direct generalization of the condensation of Abelian quasiparticles, which has been very successful to describe  the Abelian  LLL QH states.

We now discuss the bosonic state at $\nu=4/3$,  before proceeding to the general case. 
Putting $p=1$ and evaluating the correlator, \eqref{multiqp} can be written as:
\begin{multline}\label{4/3}
\psi_{4/3}=\mathcal{S}\left[ \prod_{j\in I_1} \partial_j (1-1)^2(2-2)^2(1-2) 
\right.\\ \times \left.
\prod_{j\in I_3} \partial_j(3-3)^2(4-4)^2(3-4) \right]\, ,
\end{multline}
where $(i-i)=\prod_{\alpha<\beta\in I_i} (z_\alpha-z_\beta)$ and $(i-j)=\prod_{\alpha\in I_i, \beta\in I_j}(z_\alpha- z_{\beta})$, for $i\neq j$.
This state has fundamental quasiholes with charge $e/3$ and NA statistics, even though the manifest monodromies of the operators are Abelian. This is a direct consequence of starting from a representation of the MR state, where the NA properties are coded in the Berry matrix.
Surprisingly, it is possible to find a different CFT description, where the manifest NA monodromies are recovered. The spin-polarized state \eqref{4/3} is namely closely related to the NA spin-singlet (NASS) state introduced in Ref. \onlinecite{nass}, and both are described by the same CFT. In fact, \eqref{4/3} can be rewritten as
\begin{align}
\Psi_{4/3}&=\mathcal{S} \left[\langle \prod_{j=1}^{N/2} \partial_j V_+(z_j)\prod_{j=N/2+1}^N V_-(z_j)\rangle\right]
\end{align}
with  $V_+=\psi_1(z) e^{i\sqrt{\frac{3}{4}}\varphi_c(z)+\frac{i}{2}\varphi_s(z)}$ and $V_-= \psi_2(z)e^{i\sqrt{\frac{3}{4}}\varphi_c(z)-\frac{i}{2}\varphi_s(z)}$, with $\psi_1$ and $\psi_2$ being Gepner parafermions of $su(3)_2/[u(1)]^2$.
This differs from the NASS state only by the presence of the derivatives and the symmetrizer. 
 Note that the derivatives, which originated  from the quasiparticle operators, $\mathcal{P}_\pm$, are crucial for obtaining a non-zero result after symmetrization.  Here, we use the notation introduced in Ref. \onlinecite{nass}, details on the CFT and the properties of the operators can be found there.

It is now straightforward to identify the  fundamental quasiholes with electric charge $e/3$, $H_+=\sigma_\uparrow(\eta) e^{\frac{i}{\sqrt{12}}\varphi_c(\eta)+i\varphi_s(\eta)}$ and $H_-=\sigma_\downarrow(\eta) e^{\frac{i}{\sqrt{12}}\varphi_c(\eta)-i\varphi_s(\eta)}$.\footnote{This  CFT description satisfies the hypothetical criterium for a vanishing Berry phase formulated in Ref. \onlinecite{qeop}.}
Despite the symmetrization over the spin index, the two quasiholes are still distinguishable due to the derivatives in the particle operators $V_+$. Since braiding quasiholes is by construction independent of the particle coordinates, the symmetrization over the particle positions does not affect the quasihole statistics.  The only way the symmetrization could change the statistics would be by reducing the multiplicities of the multi-quasihole states. 
Numerical tests on small systems show that this does not happen for four quasiholes, and it is therefore highly unlikely to happen for larger  numbers of quasiholes. 
All considered,  there is very strong evidence that  the state \eqref{4/3} has quasiholes that obey the same $su(3)_2$ type NA statistics as the NASS state. 

The fermionic state at $\nu=4/7$ with $q=2$ is obtained by simply changing the coefficient of the charge field $\varphi_c$. For the electron operators this means $\frac{3}{\sqrt{4\cdot 3}}\varphi_c\rightarrow\frac{7}{\sqrt{4\cdot 7}}\varphi_c$, while the quasihole operators are changed according to $\frac{1}{\sqrt{4\cdot 3}}\varphi_c\rightarrow \frac{1}{\sqrt{4\cdot 7}}\varphi_c$. The other fields are kept unchanged, which  implies that the NA part of the braiding statistics is identical in both cases, only the Abelian exchange phases differ. In the following, we confine ourselves to the case $q=1$, as multiplying with a Jastrow factor cannot change the NA properties.

The case $p=1$ is  exceptional in that there is a CFT description in which the quasihole properties are manifest. This we cannot expect for $p>1$; here we  have to rely on other methods to find the quasihole properties. In our approach all the information on NA braiding is buried  in the Berry matrix, while the multiplicity of the multi-quasihole state is coded in the distinct orderings of the quasihole operators. However, the set of all distinct orderings is not linearly independent, and to find the number of linearly independent multi-quasihole wave functions is unfeasible for large numbers of quasiholes. In Ref. \onlinecite{tt_fusion} it was however shown that it is much easier to find both the quasihole multiplicity and their fusion rules, by studying the states on a thin torus--\ie a torus where one radius becomes much smaller than the magnetic length $\ell$.
While the information on braiding is lost in this limit, the one about fusion and topological degeneracy is retained, and is in one-to-one correspondence to the properties in the physical limit where both radii are large.     In the following, we will explain how to obtain the fusion rules for the fundamental quasiholes from the thin torus analysis of \eqref{4/3}, and show that we indeed recover the fusion rules expected from the explicit CFT construction. The generalization to the other states ($p>1$) is straightforward.

On the torus, the ground state is not any longer unique, and the set of  degenerate ground states becomes particularly simple on the thin torus, where $L_x/\ell\rightarrow 0$, with $L_xL_y/\ell^2$ fixed. In this limit, the single particle states in Landau gauge are localized on rings at $y_j=-2\pi j/L_x$ and thus  well-separated, so particle hopping between different orbitals is exponentially suppressed. Thus, the ground states become crystal-like "Tao-Thouless states", specified by the occupation numbers of the single-particle orbitals\cite{TT}.  For instance, the ground state sectors of \eqref{4/3} on the thin torus are given by $|202\,202\, \ldots202\rangle$ and $|211\, 211\,\ldots 211\rangle$ plus their translations along the $y$-direction, yielding in total six ground state sectors that can be labeled by their unit cells: $(202),(220),(022)$ and $(211),(121),(112)$ respectively.  Note that for the ground states there are always four particles on three consecutive orbitals, reflecting the filling fraction $4/3$.

To find these ground states, note that before symmetrization the two layers in \eqref{4/3} are completely decoupled, with an Abelian QH state in each layer. Therefore,  the ground states of the two-layer system are given by combinations of the ground states of each layer\cite{seiyang}. Remembering that we can translate each layer independently, this yields nine ground states in the case of distinguishable particles. The final symmetrization only renders the particles indistinguishable, which reduces the number of distinct ground states, without changing the occupation numbers.  An analogous analysis yields all ground state sectors for the states with $p>1$, given by $2[02]_j[11]_{2p-1-j}$ plus translations, where $[02]_j$ means that  $02$ is repeated $j$ times.
 
Quasiholes are domain walls between different ground state sectors. For the following analysis, the relevant excitations are the fundamental quasiholes, which carry minimal charge. 
For the state \eqref{4/3} a fundamental charge $e/3$ quasihole is given by  a single string of three consecutive sites containing only three particles\cite{susch}. For instance, a domain wall between $(202)$ and $(112)$ yields such a fundamental quasihole, $2\underline{ 02\,1}12$, where the quasihole string is underlined. 
Creating a domain wall from a specific pattern is in general not unique. Starting with \eg $(211)$, one can either create a domain wall by connecting to $(202)$ or to $(121)$. This freedom reflects the presence of  several fusion channels in the CFT.  

As explained in \cite{tt_fusion}, there is a one-to-one mapping from the ground states on the thin torus to the primary fields of some CFT. Creating a domain wall with minimal charge amounts to fusing these primary fields with a fundamental quasihole. Thus finding all nonequivalent domain walls, directly yields the fusion products with the fundamental quasihole. 
In case of the state at  $\nu=4/3$, one can identify the ground states with the six primary fields of $su(3)_2$ as  $(202)={\bf 1}, (112)={\bf 3}, (022)={\bf 6}, (211)=\bar{{\bf 3}}, (121)={\bf 8}$ and $(220)=\bar{{\bf 6}}$, and the fundamental quasihole corresponds to ${\bf 3}$. 
For instance, starting with the pattern corresponding to $\bf 8$, there are two possible domain walls: 
$|\ldots\,121\,\underline{120}\,220\,\ldots\rangle$ and $|\ldots \,12\underline{1\,11}2\,112\,\ldots\rangle$,
 yielding the fusion rule ${\bf 8}\times {\bf 3}=\bar{\bf 6}+{\bf 3}$.\footnote{Note that the multiplication rules in the affine Lie algebra $su(3)_2$ differ from those in $su(3)$.}
The same analysis as for $4/3$ shows that the domain wall structure for $p>1$ is consistent with the $\mathbb{Z}_{4p-1}$ parafermion fusion rules. 
 
 To conclude, we have shown how to construct new hierarchical states by condensing genuinely NA quasiparticles. Even though we only discussed condensates on top of the MR state, this method generalizes naturally to other NA states, such as the RR states. We have also provided strong evidence that these condensate state have quasiholes obeying $\mathbb{Z}_{4p-1}$ parafermion type fusion rules, and thus have a richer NA structure than the MR parent state, which moreover is emergent from the condensation without any additional input.  In the most interesting case, the bosonic $\nu=4/3$ and fermionic $\nu=4/7$ states, we identified the relevant CFT description, where the NA properties are manifest. 
Note that the condensation procedure can be repeated, using the daughter states \eqref{NAcond} as the new parent states. This yields a hierarchy of quasiparticle condensates, similar to the one for the Abelian QH states in the LLL, albeit with a richer, NA structure.

\noindent {\bf Acknowledgements} \\
I thank Mats Horsdal, Anders Karlhede, Susanne Viefers, and especially Eddy Ardonne and Hans Hansson. 
This work was supported by NordForsk.

\end{document}